\documentclass[11pt,a4paper]{article}
\usepackage[utf8]{inputenc}
\usepackage[T1]{fontenc}
\usepackage[british]{babel}
\usepackage{amsmath, amsfonts, mathtools, amssymb, graphicx, float, lipsum, xcolor, hyperref, authblk, multicol, framed, amsthm}
\usepackage{algorithm, algpseudocode}
\hypersetup{hidelinks}
\usepackage[left=2.00cm, right=2.00cm, top=2.00cm, bottom=2.00cm]{geometry}
\graphicspath{{./figures/}}

\usepackage{array}
\usepackage{booktabs}
\usepackage{cite}
\usepackage{tikz}
\usetikzlibrary{positioning,arrows.meta,fit,calc,decorations.pathreplacing}
\tikzset{
  tx/.style={draw, rounded corners=2pt, align=center, font=\tiny\ttfamily,
             minimum height=5.5mm, inner xsep=2.5pt, inner ysep=1.5pt},
  setupb/.style={tx, fill=green!18},
  al/.style={tx, fill=violet!18},
  bo/.style={tx, fill=magenta!13},
  wt/.style={tx, fill=gray!16},
  txout/.style={tx, rounded corners=0pt, dashed, draw=black!55,
               fill=none},
  dead/.style={tx, fill=gray!8, draw=gray!60, text=gray!60},
  lbl/.style={font=\tiny, inner sep=1pt, align=center},
  note/.style={font=\tiny, align=left},
  fl/.style={-{Stealth[length=3.5pt]}, thin},
  flx/.style={-{Stealth[length=3.5pt]}, thin, gray!55},
  bus/.style={gray!75, thin},
  ref/.style={-{Stealth[length=3.5pt]}, thin, dashed, gray!70}
}
\newcommand{\hopr}{arc (180:0:0.1)}
\newcommand{\hopl}{arc (0:180:0.1)}

\usepackage{float}
\usepackage{tabularx}
\usepackage{adjustbox}
\usepackage{longtable}
\usepackage{rotating}

\title{OTS-PC: OTS-based Payment Channels for the Lightning Network}
\author[1,2]{Sergio Demian Lerner\thanks{\href{mailto:sergio@fairgate.io}{sergio@fairgate.io}}\hspace{3pt}} 

\author[1]{Ariel Futoransky\thanks{\href{mailto:futo@fairgate.io}{futo@fairgate.io}}\hspace{3pt}} 
\affil[1]{Fairgate Labs} 
\affil[2]{Rootstock Labs}
\date{}

\usepackage{xspace}

\newcommand{\txname}[1]{\texttt{#1}}
\newcommand{\ioname}[1]{\texttt{#1}}
\newcommand{\opc}[1]{\texttt{#1}} 

\newcommand{\tuple}[1]{\(\langle\) #1 \(\rangle\)}

\begin{document}
\maketitle

\begin{abstract}
We present a new type of bidirectional payment channel based on One-Time Signatures on state sequence numbers. This new construction is simpler than the Poon-Dryja construction, but provides a number of benefits such as $O(1)$ storage per channel, minimal information leakage, and compatibility with Lightning Network routing.

\end{abstract}





\section{Introduction}

Since its inception in 2009, Bitcoin~\cite{nakamoto2008bitcoin} has demonstrated that decentralized consensus can support a global monetary system without trusted intermediaries. However, the protocol’s throughput and latency limitations—roughly seven transactions per second and ten-minute block intervals—pose fundamental constraints on scalability. These parameters are essential to preserve decentralization and security, yet they make Bitcoin unsuitable for high-frequency payments or micro-transactions if used directly on-chain.

To overcome these limitations, researchers proposed off-chain payment channels, which allow two parties to conduct multiple transactions without broadcasting each to the blockchain. Only opening and closing the channel require on-chain transactions, while intermediate state updates are exchanged privately and secured by Bitcoin’s scripting primitives. This concept underpins the Lightning Network~\cite{poon2016bitcoin}, a second-layer protocol that links payment channels to a routed network, allowing instant low-fee payments between participants who do not need to share direct channels.

The Lightning Network has proven that layer two constructions can extend Bitcoin’s scalability by several orders of magnitude, while preserving its security guarantees through cryptographic enforcement and dispute mechanisms. However, current payment channel architectures still face limitations: they depend on active monitoring (watchtowers~\cite{dryja2016unlinkable,pisa}), require prefunded liquidity, and involve complex time-locked contracts that constrain usability and capital efficiency.

The notion of payment channels in Bitcoin dates even back to the early versions of the protocol: In the original implementation, Satoshi Nakamoto allowed unconfirmed transactions to be replaced by new versions if all input sequence numbers (nSequence) were increased, allowing multiple interim state changes until a final transaction was broadcast~\cite{nakamoto2010_nsequence}.
Payment channels evolved from the initial ideas, and the Poon-Dryja~\cite{poon2016bitcoin} design, now standardized in the BOLT specification~\cite{bolts}, powers the Lightning Network. Improvements continued with new protocols such as Eltoo~\cite{decker2018eltoo},   Outpost~\cite{outpost} or Sleepy~\cite{sleepy}, further reducing the requirements on watchtowers. The latest protocols achieve bidirectional payments, provide HTLCs, infinite-updates and lightweight watchtowers with acceptable privacy. However, Lightning Network channels do not provide the best combination of features: they require peers to store $O(log(N))$ hashes for $N$ channel updates, but watchtowers still need to store $O(N)$ data. In contrast, Outpost and our proposed protocol require only $O(1)$ storage for peers and watchtowers.

From an engineering and practical security point of view, the state of the payment channels is unsatisfactory: the Poon-Dryja design is complex, as is its number of transactions, protocol states, and state transitions. Historically, numerous bugs, incompatibilities with BOLT standards, and security vulnerabilities have been discovered in lightning channel implementations (for example, CVE-2020-26895 in LND). A simpler payment channel design that provides all existing features is enormously valuable. 

In 2018, Osuntokun presented the idea of \textit{Signed Sequence Numbers (SSN)} ~\cite{osuntokun2018hardeninglightning}, a method to reduce the complexity of state revocations and state storage requirements for Lightning Network nodes, using aggregated ECDSA signatures of sequence number commitments embedded in unilateral exit transactions. To the best of our knowledge, Osuntokun's SSN scheme has not been formalized. 

In this work, we present OTS Payment Channels (OTS-PCs). OTS-PCs solve the above-mentioned problems and provide the same benefits of SSNs by using hash-based One-Time Signatures. One-time signatures are possible in Bitcoin using the \txname{OP\_HASH} opcode (or any of the other hashing opcodes). Unlike SSNs, our method does not require embedded commitments.  

The core concept behind OTS-PCs is that each new state has a monotonically increasing sequence number, and the issuer of each state-committing transaction must sign the state sequence number with his OT private key before a payout transaction can be issued in the blockchain to distribute the funds. Once a party signs a state sequence number, the other party is given a window to challenge: if the signed sequence number is not the last, the other party uses a revocation path of the transaction graph to punish the first party by collecting all the funds in the channel.

To make OTS-PCs compatible with the Lightning Network, we show how OTS-PCs can support Hash-Time Locked Contracts (HTLC). We demonstrate how we can build lightweight participants and privacy-preserving watchtowers that only need to store a constant amount of information per payment channel. However, to maintain payment privacy, we do not provide a protocol for watchtowers to monitor HTLC timelocks after a successful unilateral exit.

OTS-PC channels can be extended to multi-party channels (including channel factories) easily because they have a high amount of symmetry. However, this extension is out of the scope of our work.

\section{Specification}

An OTS-PC is a protocol that allows two parties (Alice and Bob) to manage a payment channel between them. We call Alice and Bob the \textit{owners} of the channel. The payment channel resembles a shared account where the owners can dynamically re-balance the funds between them. Both owners initially deposit bitcoins in the channel, generally, but not necessarily, in equal amounts. Optional auxiliary parties, $W_a$ and $W_b$, act as watchtowers for Alice and Bob, respectively. The payment channel is defined by 3 protocols: setup, state update, and close.

An OTS-PC \textit{channel state} is defined as a funds balance between Alice and Bob at a certain time, and all open HTLC contracts with their properties (more on HTLCs later). Each channel state has a sequence number $i$, which increments (by one or more) with every channel state update. When the owners agree to move to the next state (e.g., rebalance the shared account), they must co-sign multiple connected transactions. To sign the transactions, the parties can use a 2-out-of-2 multisig or the MuSig2~\cite{musig2} scheme. MuSig2 is preferred, as it reduces the on-chain footprint and minimizes information leakage from published transactions. The owners also revoke the previous channel state by creating and signing a set of punishment transactions. 

The protocol state should not be confused with the channel state. The protocol goes through many protocol state transitions to change the channel state. Protocol participants can perform one of two actions: issue a transaction on the blockchain or send a message directly to another actor (offchain). To send and receive messages offchain, the two owners establish a peer-to-peer authenticated connection. Although the protocol defines which actions each participant can take in every state of the protocol, the publication of a transaction on the blockchain is an asynchronous event. The protocol must consider blockchain events in all states. Therefore, OTS-PC protocol state changes depending on the messages received (which can be informative or just ACKs from a previous message sent) and the transactions issued on the blockchain. Note that the protocol state is not fully determined by onchain transactions.

When one or both channel owners want to distribute the committed funds, they can close the channel. A channel can be closed cooperatively or unilaterally. A cooperative closure requires the owners to simply sign a transaction and wait for this transaction to be mined.

A unilateral close occurs when the other party is irresponsible to a party's close request. Let us assume that Alice wants to close the channel and Bob refuses to.
To unilaterally close the channel, Alice does so by signing the sequence number $i$, using a one-time signature (OTS) scheme and her one-time private key. This sequence number defines the state that Alice claims should determine the final balance distribution between her and Bob. 

If Alice uses an outdated state — say she signs $j$, with $j < i$ — a dispute mechanism is triggered to penalize her. The mechanism is based on issuing onchain two previously cosigned punishment transactions (a.k.a. revocation of old states). The first is a commitment to punish and the second is the punishment itself. The commitment transaction has an connector output that, to be spent, requires presenting as a witness a revoked sequence number signed by Alice with her OTS key. Therefore, Bob simply needs to present the OTS-signed value $j$ previously signed by Alice.  A Bitcoin script verifies the one-time signature of $j$ and checks whether $j < i$, for a value $i$ that is embedded in the Bitcoin script.  If the $j < i$ condition holds, the punishment transaction is valid and Bob can claim the funds in the channel, including all payments affected by HTLCs.

\subsection{Transaction DAG}
\label{dag}

Figure ~\ref{fig:OTS-PC} illustrates the directed acyclic graph (DAG) of transactions created by the protocol. Some of these transactions are recreated at each channel state update. Only the \txname{Setup} transaction is always published on-chain; other transactions are published only as needed.

\begin{itemize}
    \item Violet boxes: Transactions published by Alice
    \item Pink boxes: Transactions published by Bob
    \item Boxes marked $(i)$: re-created with every state update
    \item Green box: The initial funding transaction that either owner can issue.
\end{itemize}

\begin{figure}[h]
\centering
\begin{tikzpicture}[x=1cm, y=1cm]
\node[setupb, minimum width=133mm, anchor=north west] (setup) at (-1.5,0) {Setup};

\node[al, anchor=north] (ace) at (1.1,-1.35) {AliceCommitExit($i$)};
\node[bo, anchor=north] (bce) at (4.5,-1.35) {BobCommitExit($i$)};
\node[bo, anchor=north] (cpa) at (8.0,-1.35) {CommitPunishAlice($i$)};
\node[al, anchor=north] (cpb) at (11.2,-1.35) {CommitPunishBob($i$)};

\node[al, anchor=north] (aas) at (1.1,-2.75) {AliceAssertExitState($i$)};
\node[bo, anchor=north] (bas) at (4.5,-2.75) {BobAssertExitState($i$)};

\node[al, anchor=north] (afe) at (1.1,-4.15) {AliceFinalizeExit($i$)};
\node[bo, anchor=north] (bfe) at (4.5,-4.15) {BobFinalizeExit($i$)};
\node[bo, anchor=north] (pa)  at (8.0,-4.15) {PunishAlice($i$)};
\node[al, anchor=north] (pb)  at (11.2,-4.15) {PunishBob($i$)};

\node[txout, anchor=north] (ifo) at (-0.9,-2.0) {Initial\_\\Funds\_out};

\node[lbl, anchor=west] at (1.18,-0.95)  {Alice\_Exit\_out};
\node[lbl, anchor=west] at (4.58,-0.95)  {Bob\_Exit\_out};
\node[lbl, anchor=west] at (8.08,-0.95)  {Bob\_Disputes\_out};
\node[lbl, anchor=west] at (11.28,-0.95) {Alice\_Disputes\_out};

\draw[fl] (1.1,-0.62)  -- (ace.north);
\draw[fl] (4.5,-0.62)  -- (bce.north);
\draw[fl] (8.0,-0.62)  -- (cpa.north);
\draw[fl] (11.2,-0.62) -- (cpb.north);
\draw[fl] (-0.9,-0.62) -- (ifo.north);
\draw[fl] (ace) -- (aas);  \draw[fl] (bce) -- (bas);
\draw[fl] (aas) -- (afe);  \draw[fl] (bas) -- (bfe);
\draw[fl] (cpa) -- (pa);   \draw[fl] (cpb) -- (pb);

\coordinate (d1) at ($(afe.north west)+(0.40,0)$);
\coordinate (d2) at ($(bfe.north west)+(0.40,0)$);
\coordinate (d3) at ($(pa.north west)+(0.40,0)$);
\coordinate (d4) at ($(pb.north west)+(0.40,0)$);
\coordinate (b1) at (d1 |- 0,-3.70);
\coordinate (b2) at (d2 |- 0,-3.70);
\coordinate (b3) at (d3 |- 0,-3.70);
\coordinate (b4) at (d4 |- 0,-3.70);
\draw[bus] (ifo.south) -- (-0.9,-3.70) -- (b1) -- (1.0,-3.70) \hopr
           -- (b2) -- (4.4,-3.70) \hopr -- (b3) -- (7.9,-3.70) \hopr -- (b4);
\draw[fl] (b1) -- (d1);  \draw[fl] (b2) -- (d2);
\draw[fl] (b3) -- (d3);  \draw[fl] (b4) -- (d4);
\end{tikzpicture}
\caption{The transaction Direct Acyclic Graph (DAG) for an OTS-PC. Each owner has an independent exit path. All four bottom transactions spend \ioname{Initial\_Funds\_out}, so exactly one of them ever confirms. Throughout this paper a transaction is a rounded filled box and an output is a dashed rectangle.}
\label{fig:OTS-PC}
\end{figure}

Each owner has its own exit path. Alice exits by publishing \txname{AliceCommitExit($i$)}, committing to the sequence number $i$, and then \txname{AliceAssertExitState($i$)}, which carries her one-time signature over $i$. Bob's path is the mirror image, rooted at a different \txname{Setup} output and verified against his own one-time key.

The two paths are independent in both directions. Neither owner needs the other to make progress, so no owner can stall the other's exit, and the protocol needs no timeout mechanism to recover from a stalled counterparty. If Alice publishes \txname{AliceCommitExit($j$)} and then stops responding, Bob simply runs his own path to completion and the channel closes at the state he asserts.

Publishing a \txname{CommitExit} for a stale $j < i$ reveals no one-time signature and is therefore not punishable. It is also not profitable: the only way to make use of it is to publish the matching \txname{AssertExitState($j$)}, and doing so reveals a one-time signature over $j$, which lets the counterparty punish and take the whole channel. The only thing a stale \txname{CommitExit} costs is the $3\epsilon$ of the exiter's own exit output, which it consumes.

The independence of the two paths rests on an asymmetry in the signature exchange. Although both owners sign every transaction, the signature on a path's first transaction travels in one direction only: Bob sends Alice his signature on \txname{AliceCommitExit($i$)} and never receives hers, while Alice sends Bob her signature on \txname{BobCommitExit($i$)} and never receives his. If signatures on both paths were exchanged in full, an owner could publish the counterparty's \txname{CommitExit} for a stale state, consuming the counterparty's exit output and leaving them without a usable path.

In appendix \ref{formalization} we describe each transaction of the DAG.

\subsection{Setup}

\begin{itemize}
    \item Each owner chooses a hidden preimage ($P_a$ by Alice and $P_b$ by Bob) and sends the other the corresponding hash $H_a$ and $H_b$ (such that $H_a = H(P_a)$ and $H_b = H(P_b)$). These preimages are used only to let the owners agree to skip the dispute window when closing cooperatively, as described in section \ref{closing}.

    \item Each owner generates an OTS key pair: ($K^A_{priv},K^A_{pub}$) for Alice and ($K^B_{priv},K^B_{pub}$) for Bob. These keys are used to sign a sequence value $i$ in the case of unilateral exit. 
    
    \item The public keys $K^A_{pub}$  and $K^B_{pub}$ are exchanged.
 
    \item Both parties construct the \txname{Setup} transaction (a.k.a. 'Funding transaction') by providing the UTXOs containing the initial funds, but they do \emph{not} sign its inputs yet. Because SegWit transaction identifiers do not commit to witness data, the \txname{Setup} transaction identifier is already determined at this point, and both exit paths can be built against it.

    \item Both parties create the initial state-committing transactions of both exit paths:
    \begin{itemize}
        \item \txname{AliceCommitExit(0)}, \txname{AliceAssertExitState(0)}, \txname{AliceFinalizeExit(0)},
        \item \txname{BobCommitExit(0)}, \txname{BobAssertExitState(0)}, \txname{BobFinalizeExit(0)}.
    \end{itemize}
    
    These define two independent exit paths through which each owner can withdraw their initial deposit. At this point, no revocation messages exist, so challenges are not possible.

    \item Both parties sign these transactions, and exchange the signatures \emph{asymmetrically}: Bob sends Alice his signatures on Alice's path, Alice sends Bob her signatures on Bob's path, and neither returns the signatures for the counterparty's path. Only once each owner holds a fully signed path of their own do they sign the \txname{Setup} inputs and issue it on the blockchain.
\end{itemize}

The ordering of the last two steps is essential. \ioname{Initial\_Funds\_out} is a \opc{CovAddress} and therefore requires both signatures to spend. If \txname{Setup} were published before the exit transactions were signed, a party who stopped responding at that moment would leave the channel balance permanently unspendable, with no unilateral exit available to either owner. This is the same hazard that the Lightning Network avoids by exchanging \texttt{funding\_created} and \texttt{funding\_signed} before the funding transaction is broadcast.

The asymmetry of the signature exchange is equally essential, for the reason given in section \ref{dag}: an owner holding both signatures on the counterparty's \txname{CommitExit} could consume the counterparty's exit output with a stale state and leave them with no path out.

The one-time public keys $K^A_{pub}$ and $K^B_{pub}$ are shown in figures as $KA$ and $KB$.
The parties can decide the allowed range for encoding and signing $i$, but the range is bounded by Bitcoin Script rather than by the width of the chosen type. The comparison \opc{LESSTHAN} in \ioname{PunishAlice\_out} and \ioname{PunishBob\_out} operates on \texttt{CScriptNum}, which accepts at most four bytes and interprets them as a signed value, so the largest usable sequence number is $2^{31}-1$. Assuming 10 updates per second, this represents 2486 days of continuous payments, that is $2486/365 = 6.8$ years.

Implementations must reject any sequence number above $2^{31}-1$ rather than truncate or widen it. A four-byte encoding of $2^{31}$ is the non-minimal representation of negative zero and is rejected by the minimal-encoding rule, while its minimal five-byte encoding exceeds the \texttt{CScriptNum} limit for arithmetic. Either way the punishment script becomes unexecutable, so a cheater's stale exit could no longer be punished.

\subsection{Committing to a New State}

When a payment occurs (e.g., Alice pays Bob), both parties must advance to a new state and revoke the previous one. Without loss of generality, let us assume that Alice is paying Bob an amount $a$, and less assume that the payment does not create more HTLCs. Bob has a monetary incentive to complete the protocol equal to $a + z_b$. The value $z_b$ represents Bob's cost of closing the channel and opening a new one. Alice's incentive to collaborate on the update is just $z_a$, her cost for closing and opening a new channel. Because, in general $z_a \approx z_b$, Bob has a higher incentive to move to the new channel state.

To simplify the explanation, let us assume that the parties have already agreed on the amount $a$ to be transacted, and all pre-existing HTLCs (if any) are closed in the new channel update.

We also group the transactions into sets for simplicity.  The set \txname{CommitPunishX(i)} contains all commit-to-punish transactions: 
\begin{itemize}
    \item \txname{CommitPunishBob(i)},
    \item \txname{CommitPunishAlice(i)},
    \item \txname{WTA-CommitPunishBob(i)},
    \item \txname{WTB-CommitPunishAlice(i)}.
\end{itemize}  

The \txname{PunishX(i)} set contains all the punish transactions: 
\begin{itemize}
    \item \txname{PunishBob(i)}, 
    \item \txname{PunishAlice(i)},
    \item \txname{WTA-PunishBob(i)},
    \item \txname{WTB-PunishAlice(i)}.
\end{itemize}
With two independent exit paths, "party $X$ can use state $i$" means exactly one thing: $X$ holds the counterparty's signatures on $X$'s own path at $i$. A state update therefore reduces to three messages. Without loss of generality Alice is paying Bob, so Alice prefers the old state $i-1$ and Bob prefers the new state $i$.

\begin{enumerate}
    \item These transactions are built by both parties, but kept unsigned: \txname{AliceCommitExit(i)}, \txname{AliceAssertExitState(i)}, \txname{AliceFinalizeExit(i)}, their three counterparts on Bob's path, and the \txname{CommitPunishX(i)} and \txname{PunishX(i)} sets. Watchtower-related transactions are only created if watchtowers are used.

    \item The \txname{PunishX(i)} set is signed by both parties and the signatures exchanged. These are inert until the matching \txname{CommitPunishX(i)} is published, so exchanging them early is safe.

    \item \textbf{Bob $\rightarrow$ Alice}: Bob's signatures on Alice's path at $i$. Alice can now use $i-1$ or $i$; Bob can still only use $i-1$. Alice prefers $i-1$, which is the state both already hold, so Bob has given away an option Alice has no reason to take.

    \item \textbf{Alice $\rightarrow$ Bob}: \txname{CommitPunishAlice(i)} (and \txname{WTB-CommitPunishAlice(i)}) \emph{together with} Alice's signatures on Bob's path at $i$. Alice gives up $i-1$ at the same instant Bob gains $i$. Bob prefers $i$. If instead Bob were to close at $i-1$, Alice would receive more than she agreed to, so she is not exposed by his delay.

    \item \textbf{Bob $\rightarrow$ Alice}: \txname{CommitPunishBob(i)} (and \txname{WTA-CommitPunishBob(i)}). Both owners are now committed to $i$, and neither can use an earlier state without being punished.
\end{enumerate}

The two items of step 4 must not be split in the other order. Were Alice to send her signatures on Bob's path before committing herself, both owners could use the state each prefers at the same time: Alice could still close at $i-1$ and Bob could close at $i$, and whichever transaction confirmed first would decide whether the payment happened. Sending them in one message removes the window; sending \txname{CommitPunishAlice(i)} first is also safe, merely one message less efficient.

The rule behind this ordering is not that the payer commits first. It is that \emph{the party whose claim on the channel decreases must commit to the new state before the counterparty is given the ability to use it}. For a simple payment that party is the payer. When an HTLC is offered or settles, it is still the payer. But when an HTLC fails back to its sender, the party whose claim decreases is the \emph{receiver}, who loses the opportunity to claim with the preimage, and the order of steps 3 to 5 must be swapped accordingly. An implementation that always makes the payer commit first would open exactly the race described above on every HTLC failure.

Finally, each owner signs the counterparty's punishment transactions destined for their own watchtower and forwards them, as described in section \ref{watchtowers}.

\subsection{Example}

In figure ~\ref{fig:OTS-PC-EXAMPLE} we show an example of an onchain dispute where Alice tries to cheat by attempting a unilateral exit with an old state $j=50$, while the latest agreed state is $i=100$. Alice publishes \txname{AliceCommitExit(50)}, which by itself reveals nothing and cannot be punished, and then \txname{AliceAssertExitState(50)}, whose witness carries her one-time signature over $50$. That signature is what Bob needs. He publishes \txname{CommitPunishAlice(100)} and \txname{PunishAlice(100)}, the second of which spends \ioname{Initial\_Funds\_out} and pays him the whole channel. Neither punishment transaction is timelocked, so Bob has the entire dispute window to confirm both of them, while Alice's \txname{AliceFinalizeExit(50)} cannot be published until that window has elapsed. Once \ioname{Initial\_Funds\_out} is spent by the punishment, Alice's exit is dead.

\begin{figure}[h]
\centering
\begin{tikzpicture}[x=1cm, y=1cm]
\node[al, anchor=north] (ac) at (0,0)      {AliceCommitExit(50)};
\node[al, anchor=north] (aa) at (0,-1.1)   {AliceAssertExitState(50)};
\node[dead, anchor=north] (af) at (0,-2.6) {AliceFinalizeExit(50)};

\node[bo, anchor=north] (cp) at (5.4,-1.1) {CommitPunishAlice(100)};
\node[bo, anchor=north] (pa) at (5.4,-2.6) {PunishAlice(100)};

\node[txout, anchor=north] (ifo) at (8.6,0) {Initial\_\\Funds\_out};

\draw[fl] (ac) -- (aa);
\draw[flx] (aa) -- node[note, right=1mm, pos=0.3, text=gray!70] {blocked: needs the\\dispute window} (af);
\draw[fl] (cp) -- (pa);
\draw[ref] (aa.east) -- node[note, above, midway] {reveals $OTS_A(50)$} (cp.west);

\coordinate (dl) at ($(af.north west)+(0.40,0)$);
\coordinate (dr) at ($(pa.north east)-(0.40,0)$);
\coordinate (bl) at (dl |- 0,-2.35);
\coordinate (br) at (dr |- 0,-2.35);
\draw[bus] (ifo.south) -- (8.6,-2.35);
\draw[bus] (8.6,-2.35) -- node[note, above] {spent here first} (br);
\draw[fl]  (br) -- (dr);
\draw[bus] (br) -- (5.5,-2.35) \hopl -- (0.1,-2.35) \hopl -- (bl);
\draw[flx] (bl) -- (dl);
\end{tikzpicture}
\caption{The transactions involved in a dispute. Alice tries to unilaterally exit with state $j=50$, but Bob punishes her using the punishment transactions created in state $i=100$. Greyed transactions never confirm.}
\label{fig:OTS-PC-EXAMPLE}
\end{figure}

\subsection{Liveness Guarantees During Unilateral Exits}

Because each owner holds an exit path that the other cannot consume, stalling is not a threat and the protocol needs no timeout mechanism. If Alice publishes \txname{AliceCommitExit($j$)} and then refuses to continue, Bob does not need to force her forward or to time her out: he publishes his own \txname{BobCommitExit($i$)} and \txname{BobAssertExitState($i$)} and closes the channel at the state he asserts. Alice's abandoned path consumes only her own $3\epsilon$ exit output.

Nor does Bob gain anything by publishing \txname{BobAssertExitState($j$)} for a stale $j$ in order to resolve Alice's exit for her, as he would reveal a one-time signature over $j$ and be punished. Each owner asserts only the latest state, using only their own key.

\subsection{Symmetric vs Asymmetric Exists}

In the Lightning Network payment channel specification, if a party  (e.g. Alice) attempts a unilateral exit based on a state $j$, she gives Bob immediate access to his share of the funds according to the balances associated with the state $j$, but Alice does not get immediate access to her share. This is known as an asymmetric exit. The OTS-PCs design presented here does not attempt such an asymmetry. Symmetrical exit is not only more fair, but is also simpler to implement. 

It is worth being precise about what is symmetric here, because the construction of section \ref{dag} already gives each owner a separate exit path. \emph{Structure} and \emph{payout} are independent choices. Our two paths exist so that neither owner can stall the other, and both of them pay out identically: \txname{AliceFinalizeExit($i$)} and \txname{BobFinalizeExit($i$)} each pay $A_{BAL}(i)$ to Alice and $B_{BAL}(i)$ to Bob, and neither imposes any delay on the party that did not initiate the exit. The exit is therefore symmetric in payout regardless of which path is used.

An asymmetric variant would keep the same two paths and change what they pay: the path used by party $X$ would give the other party $Y$ immediate access to his portion of the funds while delaying $X$'s own. Figure ~\ref{fig:OTS-PC-ASYNC} depicts this design variation. We do not adopt it, for the reasons that follow.

\begin{figure}[h]
\centering
\begin{tikzpicture}[x=1cm, y=1cm]
\node[setupb, minimum width=101mm, anchor=north west] (su) at (-0.5,0) {Setup};
\node[lbl, anchor=west] at (1.28,-0.92) {Initial\_Funds\_out};
\node[lbl, anchor=west] at (6.68,-0.92) {Alice\_Exit\_out};

\node[txout, anchor=north] (ifo) at (1.2,-1.25) {Initial\_Funds\_out};
\node[al, anchor=north]    (ace) at (6.6,-1.25) {AliceCommitExit($i$)};
\node[al, anchor=north]    (aas) at (3.9,-2.70) {AliceAssertExitState($i$)};

\node[txout, anchor=north] (bs)  at (1.0,-4.05) {$B_{BAL}(i)$ to Bob};
\node[txout, anchor=north] (rem) at (5.4,-4.05) {AliceRemainder\_out($i$)};
\node[al, anchor=north] (af)  at (3.5,-5.35) {AliceFinalizeExit($i$)};
\node[bo, anchor=north] (pun) at (7.8,-5.35) {PunishAlice($i$,$j$)};
\node[txout, anchor=north] (ao)  at (3.5,-6.68) {$A_{BAL}(i)$ to Alice};

\node[lbl, anchor=north, text=gray!60] at (1.0,-4.75) {immediate};
\node[lbl, anchor=west,  text=gray!60] at (3.6,-6.15) {after timelock};
\node[lbl, anchor=north, text=red!65]  at (7.8,-6.02) {one per $(i,j)$ pair};

\draw[fl] (1.2,-0.62) -- (ifo.north);
\draw[fl] (6.6,-0.62) -- (ace.north);

\draw[fl] (ifo.south) -- (1.2,-2.25) -- (2.75,-2.25) -- (2.75,-2.70);
\draw[fl] (ace.south) -- (6.6,-2.25) -- (5.05,-2.25) -- (5.05,-2.70);

\draw[fl] (2.75,-3.25) -- (2.75,-3.65) -- (1.0,-3.65) -- (1.0,-4.05);
\draw[fl] (5.05,-3.25) -- (5.05,-3.65) -- (5.4,-3.65) -- (5.4,-4.05);

\draw[fl] (5.0,-4.60) -- (5.0,-4.98) -- (3.5,-4.98) -- (3.5,-5.35);
\draw[fl, red!55] (5.8,-4.60) -- (5.8,-4.98) -- (7.8,-4.98) -- (7.8,-5.35);

\draw[fl] (af) -- (ao);

\node[note, anchor=north west, text width=100mm] at (-0.5,-7.35)
  {Alice's exit is shown; Bob's is the mirror image. To pay Bob immediately, \txname{AliceAssertExitState($i$)} must spend \ioname{Initial\_Funds\_out}, so the punishable remainder becomes an output of a per-state transaction instead of an output of \txname{Setup}. The punishment can no longer be pre-signed once per state.};
\end{tikzpicture}
\caption{The transaction Direct Acyclic Graph (DAG) for an OTC-PC with asymmetric exits}
\label{fig:OTS-PC-ASYNC}
\end{figure}

Beyond the economics discussed below, this variation also costs the construction its $O(1)$ storage, and for a structural reason. To pay $Y$ immediately, \txname{XAssertExitState($i$)} must spend \ioname{Initial\_Funds\_out}, so the punishable remainder no longer sits at an outpoint fixed by \txname{Setup}: it becomes an output of \txname{XAssertExitState($i$)}, and which $i$ that is depends on the state the cheating owner chooses to publish. A punishment transaction signed at state $i$ cannot name an outpoint that the counterparty will only create later, so a separate punishment must be pre-signed for every pair of states, and the owners must retain $O(N)$ of them rather than only the latest. In the symmetric design the comparison \opc{LESSTHAN} inside \ioname{PunishX\_out} covers every stale $j$ with a single transaction precisely because \ioname{Initial\_Funds\_out} does not move until the dispute is settled.

It seems beneficial that no party can impose a delay on the other party's access to her funds when exiting unilaterally. However, this apparent benefit does not make the scheme fairer: By just stopping communicating, one party forces the other to unilaterally exit, imposing the same delay on the other party. The fact that one party can impose a delay on the other party's ability to access her funds is inevitable, and exists in any protocol based on fraud-proofs. Therefore, the asymmetric exit does not reduce the financial cost in the worst case. Instead, under certain conditions, it creates a prisoner's dilemma in which neither party is incentivized to cooperate to close a channel. 

Table ~\ref{tab:dilemma} shows the payouts for all exit cases when both parties have approximately the same amount of funds assigned to them. We define $p$ as the opportunity cost imposed by the delay to access funds, $u$ as the fees required to pay for the transactions in the unilateral exit path, and $c$ as the fees that must be paid for a cooperative close.

\begin{table}
    \centering
    \begin{tabular}{|c|c|c|}
    \hline
         & Alice cooperates &  Alice does not cooperate\\
         \hline
       Bob Cooperates  & 
       \shortstack{Alice: $c/2$ \\ Bob: $c/2$} & \shortstack{Alice:  0 \\ Bob: $-(p+u)$} \\
       \hline
       Bob does not cooperate  & 
       \shortstack{Alice: $-(p+u)$ \\ Bob: 0} & 
       \shortstack{Alice: $-(p+u)/2$ \\ Bob: $-(p+u)/2$}  \\
       \hline
    \end{tabular}
    \caption{A Prisoner's Dilemma}
    \label{tab:dilemma}
\end{table}

As long as $(p+u) > c$, this payout table represents a prisoner's dilemma, and the rational strategy of both parties is to expect the other party to close the channel unilaterally first.

In case one party owns a very small amount of channel funds, comparable to the cooperative closing fees, then that party ceases to have any incentive to misbehave.

The best way to avoid the prisoner's dilemma is to make the unilateral exit symmetric.
Let us assume Bob is the participant that unilaterally exits. Other unsatisfactory solutions may be the following:
\begin{enumerate}

    \item In Bob's unilateral exit transactions, reduce Alice's balance in $c_b$ and reassign this value to Bob. Since the states committing transactions are created in advance, the downside of this solution is that the value $c_b$ assigned may not match the actual cost $c$ in fees at the time of exit.
    
    \item Impose a dynamic cost of $c$ on Alice to retrieve the funds.  We add to the \txname{AliceExit(i)} transaction a time-locked path that returns $c_b$ funds to Bob, where $c_b$ is chosen so that ($c_b > c$) even under blockchain congestion. Consuming Alice’s funds from the \txname{AliceExit(i)} transaction must use approximately the same number of virtual bytes as the cooperative exit transaction, so it costs approximately $c$. In this way, Alice is incentivized to transfer her funds from \txname{AliceExit(i)} quickly, so she ends up paying the cost $c$ at the time of exit and not $c_b$. The downside of this solution is that Alice is forced to react before a certain time limit, even if the unilateral exit attempt was not malicious. 
\end{enumerate}

\subsection{Closing the Channel}
\label{closing}

If both parties agree to close the channel, they co-sign a payment transaction that spends directly from the \txname{Setup} transaction, distributing the funds accordingly. If funds remain in intermediate connector outputs (used for pre-paying fees or managing dust outputs), they can also be collected and shared.

If one party becomes uncooperative, the other performs a unilateral close using their own exit path. Taking Alice as the exiting party:

\begin{enumerate}
    \item Publish \txname{AliceCommitExit($i$)} and \txname{AliceAssertExitState($i$)}.
    \item Wait for a potential dispute.
    \item Publish \txname{AliceFinalizeExit($i$)}, which is timelocked to allow dispute resolution.
\end{enumerate}

Bob does not need to participate at any point, and cannot interfere: he holds no signature that would let him publish or consume any transaction of Alice's path. Should he wish to close at the same time, he runs his own path in parallel; both paths pay the same amounts, and whichever \txname{FinalizeExit} spends \ioname{Initial\_Funds\_out} first settles the channel. The other path is left with its $\epsilon$-sized \ioname{Ready\_out} and \ioname{CPFP\_out} unspent, which is the only cost of the duplication.

In case Bob is not responding temporarily, Alice attempts a unilateral exit with the last valid state. Bob may wake up and realize that \txname{AliceCommitExit($i$)} and \txname{AliceAssertExitState($i$)} have been issued onchain. In this case, we add a happy resolution path for Bob to accelerate the \txname{AliceFinalizeExit($i$)} transaction before the dispute timeout by allowing it to be issued immediately providing both $P_a$ and $P_b$ preimages, such that $H(P_a) = H_a$ and $H(P_b) = H_b$. This is the only use the preimages have in the construction.

\subsection{Watchtowers}
\label{watchtowers}

We now examine variants of the OTS-PC scheme that influence the degree of privacy achievable by peers when employing watchtowers. Depending on the level of privacy desired, watchtowers may need their own punishment transactions or share them with one of the parties. We will first assume that these are required, but later we will discuss in which cases we can share them.

To use neutral notation, we use the prefix \txname{WTX} to identify transactions given to the watchtower $W_X$ that serves $X$. The punishment transactions used by $W_X$ to punish $Y$ are labeled \txname{WTX-CommitPunishY(i)} and \txname{WTX-PunishY(i)}.

Each party can decide what information is leaked to his watchtower, so we define different per-party privacy levels.

\subsubsection{Level 1: Balance Privacy}

Watchtowers receive periodic punishment transactions together with the transaction ID of the \txname{Setup} transaction. $X$’s watchtower receives \txname{WTX-CommitPunishY} and \txname{WTX-PunishY}. Since each owner exits through its own path, $X$'s watchtower only ever needs to monitor $Y$'s exit output, \ioname{Y\_Exit\_out}, and the \txname{YAssertExitState} transaction that reveals $Y$'s one-time signature. The watchtower branch is shown in Figure~\ref{fig:OTS-PC-WT}; the owners' own exit and punishment branches are those of Figure~\ref{fig:OTS-PC}.

\begin{figure}[h]
\centering
\begin{tikzpicture}[x=1cm, y=1cm]
\node[setupb, minimum width=92mm, anchor=north west] (su) at (-0.5,0) {Setup};
\node[lbl, anchor=west] at (1.48,-0.92) {WTB\_Disputes\_out};
\node[lbl, anchor=west] at (7.08,-0.92) {WTA\_Disputes\_out};

\node[wt, anchor=north] (cb) at (1.4,-1.25) {WTB-CommitPunishAlice($i$)};
\node[wt, anchor=north] (ca) at (7.0,-1.25) {WTA-CommitPunishBob($i$)};
\node[txout, anchor=north] (ifo) at (4.2,-1.25) {Initial\_\\Funds\_out};
\node[wt, anchor=north] (pb) at (1.4,-2.55) {WTB-PunishAlice($i$)};
\node[wt, anchor=north] (pa) at (7.0,-2.55) {WTA-PunishBob($i$)};

\draw[fl] (1.4,-0.62) -- (cb.north);
\draw[fl] (7.0,-0.62) -- (ca.north);
\draw[fl] (4.2,-0.62) -- (ifo.north);
\draw[fl] (cb) -- (pb);  \draw[fl] (ca) -- (pa);

\coordinate (dl) at ($(pb.north east)-(0.40,0)$);
\coordinate (dr) at ($(pa.north west)+(0.40,0)$);
\coordinate (bl) at (dl |- 0,-2.28);
\coordinate (br) at (dr |- 0,-2.28);
\draw[bus] (ifo.south) -- (4.2,-2.28);
\draw[bus] (4.2,-2.28) -- (bl);  \draw[fl] (bl) -- (dl);
\draw[bus] (4.2,-2.28) -- (br);  \draw[fl] (br) -- (dr);

\node[note, anchor=north west, text width=92mm] at (-0.5,-3.45)
  {Bob's watchtower WTB punishes Alice, and Alice's watchtower WTA punishes Bob. Each branch spends a dispute output fixed at \txname{Setup}, so a single pair per state suffices and the watchtower retains only the latest one. Neither transaction carries a balance, so the watchtower learns no payment amounts.};
\end{tikzpicture}
\caption{The watchtower branch under level 1 privacy}
\label{fig:OTS-PC-WT}
\end{figure}

Watchtowers learn which channels are updated, at what rate, and the total amount involved in the channel. However, watchtowers cannot learn channel balances at each state update, because the \txname{WTX-CommitPunishY(i)} and \txname{WTX-PunishY(i)} transactions do not carry the state balances and only handle misbehavior. Therefore, watchtowers cannot learn payment amounts.

Note that a party's watchtower cannot misbehave and make that party lose a dispute because the watchtower does not have the OTS private key to sign a state sequence number. 

\subsubsection{Level 2: Channel ID  Privacy}

We now show how we can increase channel privacy. Watchtowers will not be able to learn anything about a channel until a dispute is opened onchain. In case of a dispute, watchtowers learn only the total amount locked in the channel, but not the payment amounts. Figure ~\ref{fig:OTS-PC-L2} shows the changes required.

\begin{figure}[h]
\centering
\begin{tikzpicture}[x=1cm, y=1cm]
\node[setupb, minimum width=58mm, anchor=north west] (su) at (0,0) {Setup};
\node[lbl, anchor=west] at (2.98,-0.92) {WT\_Disputes\_out};
\node[setupb, anchor=north] (wd) at (2.9,-1.25) {WTDisputes \quad OP\_RETURN $\langle x\rangle$};

\node[wt, anchor=north] (wca) at (1.1,-2.6) {WTB-CommitPunishAlice($i$)};
\node[wt, anchor=north] (wcb) at (4.7,-2.6) {WTA-CommitPunishBob($i$)};
\node[wt, anchor=north] (wpa) at (1.1,-3.8) {WTB-PunishAlice($i$)};
\node[wt, anchor=north] (wpb) at (4.7,-3.8) {WTA-PunishBob($i$)};

\draw[fl] (2.9,-0.62) -- (wd.north);
\draw[fl] (2.1,-1.80) -- (2.1,-2.20) -- (1.1,-2.20) -- (1.1,-2.60);
\draw[fl] (3.7,-1.80) -- (3.7,-2.20) -- (4.7,-2.20) -- (4.7,-2.60);
\draw[fl] (wca) -- (wpa);  \draw[fl] (wcb) -- (wpb);

\node[note, anchor=north west, text width=88mm] at (0,-4.6)
  {Only the watchtower branch changes. The dispute transactions no longer hang off \txname{Setup} directly: they are attached to a fixed \txname{WTDisputes} transaction whose identifier watchtowers cannot predict, published only when a watchtower intervenes. The two exit paths and the owners' own punishment branches are exactly as in Figure \ref{fig:OTS-PC}.};
\end{tikzpicture}
\caption{The transaction Direct Acyclic Graph (DAG) for an OTS-PC with Privacy Level 2}
\label{fig:OTS-PC-L2}
\end{figure}

The dispute transactions (\txname{WTX-CommitPunishY}/\txname{WTX-PunishY}) are detached from the \txname{Setup} transaction. Instead, they are attached to a fixed \txname{WTDisputes} transaction. This transaction is only issued onchain when a watchtower intervenes and it needs to have a transaction ID that watchtowers cannot predict. It can have a \txname{OP\_RETURN} output pushing a private random number $x$ only known to Alice and Bob, or use unique randomly tweaked output addresses.

Inserting \txname{WTDisputes} lengthens the watchtower dispute chain by one transaction, so the funding changes: instead of two separate $2\epsilon$ outputs, \txname{Setup} carries a single output of $5\epsilon$ feeding \txname{WTDisputes}, which emits $2\epsilon$ for each watchtower branch and pays $\epsilon$ in fees. The owners' own dispute outputs are unaffected.

Watchtowers periodically receive an encrypted packet $W_q(j)$ for each channel $q$, where $q$ is the transaction ID of the transaction \txname{WTDisputes}. The subscript $q$ identifies the channel and is not a key; the key is introduced below. Upon receiving $W_q(j)$, the watchtower discards the data $W_q(j-1)$ previously received and stores $W_q(j)$ in its place. The packet $W_q(j)$ is encrypted with a new IV each time, so the watchtower cannot see if two packets are encryptions of the same punishment transactions or not. 

To further improve privacy, we make the following changes:

\begin{enumerate}
   
\item  The client does not send every $W_q(j)$ to the watchtower, but sends a subset of packets $W_q(j)$ ($i \in J_q$),  at fixed intervals (e.g., one every second). If the channel does not change state, then the same state is sent encrypted with a different IV.  

\item The two parties add a random offset to the sequence numbers when creating a new state to be informed to watchtowers, and also for the state update immediately following an informed state. The purpose of this second random offset is to prevent the leakage of information about the real channel update rate in the event of unilateral exit, as the sequence number is revealed by the transaction \txname{AssertExitState(i)}. 

\end{enumerate}

More formally, we define two sequences. The \textit{Internal Sequence Number (or ISN)} $j$ is monotonically incremented by 1 at each new state. The \textit{External Sequence Number (or ESN)} $S(j)$ is the value that will be one-time signed for ISN $j$. The ESN numbers $S(j)$ are chosen so that $S(j) = S(j-1) + d(j)$ for ($j>0$) and $S(0) = d(0)$. For $j \notin J_q$ and $(j+1) \notin J_q$, $d(i)= 1$. For $j \in J_q$ or $(j+1) \in J_q$ , $d(i)$ are chosen uniformly at random within an interval, so that the ESN update rate observed by the watchtower matches the maximum update rate expected for any payment channel. The ESN update rate should be standardized across all implementations to prevent watchtowers from inferring common ownership of multiple channels based on correlated update frequencies.

The watchtowers will never learn the ISN of a state, neither they can correlate events from channel updates from different clients. The channel update rate is hidden even in the case of unilateral exit. 

For the watchtower of a party $X$, the packet $W_q(j)$ is an encryption of the payload $T_q(j)$ that comprises the punishment transactions \txname{WTDisputes}, WTX-\txname{CommitPunishY(i)} and \txname{WTX-PunishY(i)}. Each \txname{CommitExit(i)} transaction forces the matching \txname{AssertExitState(i)} transaction to reveal a pre-image $P_e$ of a committed hash $H_e$. Both exit paths commit to the same $H_e$, so a watchtower can decrypt the packet whichever path its client's counterparty takes. This pre-image $P_e$ is used as the key that encrypts the package given to watchtowers, $W_q(j) = \mathrm{Enc}(P_e,\, T_q(j))$.  The watchtower client (Alice or Bob) is responsible for the correct encryption of $T_q(j)$. The client is incentivized to perform the encryption correctly. If he does not provide the correct encrypted packet, the watchtower will not be able to defend the client in the event of a dispute. 

When a watchtower observes that a \txname{CommitExit(j)} transaction is being issued, it recovers the key $P_e$, and uses it to decrypt the last packet it has received for that channel, recovering the corresponding payload $T_q$. Now the watchtower can extract the punishment transactions from $T_q(j)$ and issue the \txname{WTDisputes} transaction onchain, followed by the \txname{WTX-CommitPunishY} and \txname{WTX-PunishY} transactions. 

Note that if the watchtower stored the previous packets, it could decrypt all of them when a unilateral exit occurs. However, if the channel parties have obfuscated the update rate (by sending messages to watchtowers at regular intervals and by skipping sequence numbers between channel updates), then the watchtowers cannot learn anything related to the channel. 

\subsubsection{Level 3: Channel, Amount and Update Rate Privacy after Disputes}

To achieve greater privacy even after disputes, we use a mechanism similar to Outpost~\cite{outpost} but adapted to our scheme. We start with the design for level 2 privacy, but add a new fixed transaction \txname{StartExit} between the \txname{Setup} and \txname{CommitExit(j)} transactions. This lengthens each exit chain by one transaction, so \ioname{Alice\_Exit\_out} and \ioname{Bob\_Exit\_out} must carry $4\epsilon$ rather than $3\epsilon$. Since the two exit paths must remain independent, there is one such transaction per path, \txname{AliceStartExit} and \txname{BobStartExit}: a single shared one would have to fund both paths, and only one of them could spend it. This costs nothing in practice, because a watchtower serving $X$ only ever watches $Y$'s path and therefore only needs $Y$'s identifier; where the single-path design gave both watchtowers one shared channel identifier, each watchtower now holds exactly the one it needs. The transaction \txname{StartExit} needs to have an ID watchtowers cannot predict, so either it holds an \txname{OP\_RETURN} with a random value $x$, or Alice and Bob covenant address that connects to the \txname{CommitExit(i)} transaction is randomly changed (e.g. via public key tweaking) to be unique. We therefore write $q_A$ for the transaction ID of \txname{AliceStartExit} and $q_B$ for that of \txname{BobStartExit}. These replace the single identifier $q$ of level 2, which was the identifier of \txname{WTDisputes}. The identifier held by a watchtower is always the one of the path it watches, which is the counterparty's: Alice's watchtower holds $q_B$, and Bob's watchtower holds $q_A$. Note that the two subscripts carry different letters for the same object: the packet $W^B_{q_A}$ is used by Bob's watchtower and is filed under Alice's path identifier.

The encrypted packets follow the same logic, and are placed crosswise. $X$'s watchtower only acts when $Y$ cheats, that is, when $Y$'s path appears onchain, so the packet $W^X_{q_Y}(i+1)$ must ride in the \txname{OP\_RETURN} of \txname{YCommitExit($i$)}. Placing each packet on its owner's own path would make it unreachable: that path is published only when its owner exits, which is precisely the case in which that owner's watchtower has nothing to do.
Figure ~\ref{fig:OTS-PC-L3} presents the DAG for level 3 watchtower privacy.

\begin{figure}[h]
\centering
\begin{tikzpicture}[x=1cm, y=1cm]
\node[setupb, minimum width=104mm, anchor=north west] (su) at (0,0) {Setup};
\node[lbl, anchor=west] at (2.68,-0.92) {Alice\_Exit\_out};
\node[lbl, anchor=west] at (7.88,-0.92) {Bob\_Exit\_out};

\node[al, anchor=north, align=center] (ase) at (2.6,-1.25) {AliceStartExit\\OP\_RETURN $\langle x_A\rangle$};
\node[bo, anchor=north, align=center] (bse) at (7.8,-1.25) {BobStartExit\\OP\_RETURN $\langle x_B\rangle$};

\node[al, anchor=north, align=center] (ac) at (2.6,-2.85) {AliceCommitExit($i$)\\OP\_RETURN $\langle W^B_{q_A}(i{+}1)\rangle$};
\node[bo, anchor=north, align=center] (bc) at (7.8,-2.85) {BobCommitExit($i$)\\OP\_RETURN $\langle W^A_{q_B}(i{+}1)\rangle$};

\node[al, anchor=north] (aa) at (2.6,-4.35) {AliceAssertExitState($i$)};
\node[bo, anchor=north] (ba) at (7.8,-4.35) {BobAssertExitState($i$)};
\node[al, anchor=north] (af) at (2.6,-5.45) {AliceFinalizeExit($i$)};
\node[bo, anchor=north] (bf) at (7.8,-5.45) {BobFinalizeExit($i$)};

\draw[fl] (2.6,-0.62) -- (ase.north);  \draw[fl] (7.8,-0.62) -- (bse.north);
\draw[fl] (ase) -- (ac);  \draw[fl] (bse) -- (bc);
\draw[fl] (ac) -- (aa);   \draw[fl] (bc) -- (ba);
\draw[fl] (aa) -- (af);   \draw[fl] (ba) -- (bf);

\node[note, anchor=north, text width=40mm, text=gray!65] at (5.2,-2.0)
  {$q_A$, $q_B$ = txids of \txname{AliceStartExit} and \txname{BobStartExit}};
\node[note, anchor=north west, text width=100mm] at (0,-6.25)
  {One \txname{StartExit} per path: a shared one would have to fund both paths and only one could spend it. Each watchtower holds the identifier of the path it watches, which is the counterparty's: Alice's watchtower holds $q_B$ and Bob's holds $q_A$. The encrypted packets ride crosswise for the same reason: $W^A_{q_B}$, which Alice's watchtower needs, is carried by \emph{Bob's} \txname{CommitExit}, since that is the transaction published when Bob cheats.};
\end{tikzpicture}
\caption{The transaction Direct Acyclic Graph (DAG) for an OTS-PC with Privacy Level 3}
\label{fig:OTS-PC-L3}
\end{figure}

Without loss of generality, let us assume that only Alice is using a watchtower. The packet $W^A_{q_B}(i+1)$ for the watchtower WTA allows revocation of states $j$ $(j < i+1)$ and is stored in a transaction output containing a  \txname{OP\_RETURN}, in the \txname{CommitExit(i)} transaction on the path identified by $q_B$. The \txname{OP\_RETURN} will also contain the index $i$ in plaintext form. $T^A_{q_B}(i+1)$ will contain only the \txname{WTA-CommitPunishBob(i)} and \txname{WTA-PunishBob(i)} transactions required.

Instead of both parties using a single decryption key $P_e$, Bob creates a chain of decryption keys $K^A(i-1)=H(K^A(i))$. Note that Bob creates the key $K^A$ to encrypt Alice’s watchtower data. For the first key $K^A(M)$, we use a sufficiently high $M$. Our payment channel will support $M$ state updates before requiring closing, so $M$ is bounded by the same Script limit as the sequence numbers themselves: what is one-time signed and compared with \opc{LESSTHAN} is the external sequence number $S(j)$, which must satisfy $S(M) \leq 2^{31}-1$. Because $S(j) \geq j$, and strictly exceeds it whenever a random offset is applied, $M$ must be chosen correspondingly below $2^{31}-1$ rather than at it. When creating the state update $i$, Alice will receive from Bob the packet $W^A_{q_B}(i+1)$ encrypted with the key $K^A(i+1)$ ($W^A_{q_B}(i+1) =  E_{K^A(i+1)}( T^A_{q_B}(i+1))$. Alice will store the packet in the transaction output pushing its value after the \txname{OP\_RETURN}. Alice cannot verify the correctness of $W^A_{q_B}(i+1)$ at this point, but she continues with the protocol.

When the next channel state with sequence number $(i+1)$ is created, Bob will reveal to Alice the key $K^A(i+1)$. Alice will verify that $W^A_{q_B}(i+1)$ (embedded in the previous state update transactions) was correctly encrypted. If not, then Alice aborts the state update and unilaterally exits the channel. Although we can force the revelation of $K^A(i+1)$ to occur before Bob can use the new channel state, it is not important, since if $W^A_{q_B}(i+1)$ is incorrect, Alice (who is currently online) can immediately start a unilateral exit without the help of a watchtower. 
If $W^A_{q_B}(i+1)$ is correct, then Alice sends her watchtower the tuple ($q_B$, $K^A(i+1)$, $i+1$). The watchtower discards any previous entry related to $q_B$, and stores this new tuple. 

When a dispute occurs onchain, Bob's \txname{BobStartExit} transaction will be posted onchain (with ID $q_B$) and so watchtowers can scan their directory looking for an entry related to the identifier $q_B$ and retrieve the last tuple ($i+1$, $K^A(i+1)$). At that point, the watchtower waits until \txname{BobCommitExit} is published to recover the state update index $j$. This information is enough for the watchtower to compute $K^A(j+1)$ from $K^A(i+1)$ by chained hashing. Afterward, the watchtower can decrypt $W^A_{q_B}(j+1)$ --- the packet carried by the \txname{BobCommitExit($j$)} that was actually published, not the one for the latest state --- and obtain the \txname{WTA-CommitPunishBob($j$)} and \txname{WTA-PunishBob($j$)} transactions. Figure~\ref{fig:keychain} shows the three values involved and the hashing that links them.

\begin{figure}[h]
\centering
\begin{tikzpicture}[x=1cm, y=1cm,
  key/.style={draw, rounded corners=2pt, font=\tiny, minimum height=5mm,
              minimum width=9mm, inner sep=1pt, fill=gray!8},
  have/.style={key, fill=green!22},
  need/.style={key, fill=orange!30}]

\foreach \n in {1,...,10}{ \node[key] (k\n) at ({(\n-1)*1.18},0) {$K^A(\n)$}; }
\node[need] (k4) at ({3*1.18},0) {$K^A(4)$};
\node[have] (k9) at ({8*1.18},0) {$K^A(9)$};

\foreach \n in {1,...,9}{ \pgfmathtruncatemacro{\m}{\n+1} \draw[fl] (k\m) -- (k\n); }
\node[lbl, anchor=south] at ({2.1*1.18},0.40) {$K^A(n-1)=H(K^A(n))$, one way only};

\draw[decorate, decoration={brace, amplitude=4pt, mirror}, gray!70]
  ({3*1.18-0.45},-0.42) -- ({8*1.18+0.45},-0.42)
  node[midway, below=4pt, font=\tiny, text=gray!75] {5 hashes};

\node[lbl, anchor=south, align=center, text width=33mm] (dbn) at ({8*1.18},0.40)
  {\textbf{in its database}\\ $(i{+}1,\,K^A(i{+}1))$ with $i{+}1=9$,\\ sent by Alice at the last update};
\draw[fl, green!45!black] (dbn.south) -- (k9.north);

\node[lbl, anchor=north, align=center, text width=40mm] (chn) at ({1.4*1.18},-1.75)
  {\textbf{from the blockchain}\\ \txname{BobCommitExit($j$)} with $j=3$,\\ whose \txname{OP\_RETURN} carries $W^A_{q_B}(4)$};
\draw[fl] (chn.north) -- (k4.south);

\node[lbl, anchor=west, align=left, text width=44mm] (out) at ({8.0*1.18},-1.75)
  {\textbf{must compute}\\ $K^A(4)=H^{5}(K^A(9))$,\\ then decrypt $W^A_{q_B}(4)$ with it};
\draw[fl, orange!65!black] (k4.south east) to[out=-30, in=180] (out.west);
\end{tikzpicture}
\caption{The watchtower holds one key of the chain and hashes forward to reach the one it needs. Holding $K^A(9)$ yields every earlier key and no later one, so a watchtower can punish any stale state but learns nothing about states still to come.}
\label{fig:keychain}
\end{figure}

We can use the ShaChain trick~\cite{lightning_bolts_03_transactions} to recover $K^A(j+1)$ in $O(log(M))$ hash computations instead of $O(M)$, but store $O(log(N))$ hashes per channel instead of $O(1)$.

To make our scheme more efficient, Alice’s \txname{OP\_RETURN} output will store only the encrypted Musig2-aggregated Schnorr signatures of the \txname{WTA-CommitPunishBob(i)} and \txname{WTA-PunishBob(i)} transactions (totaling 130 bytes). 

Note that in the Lightning Network protocol the watchtower learns the final balance split at the time of channel closure, whereas in the OTS-PC scheme it learns it only in case the unilateral exit is uncontested, because the \txname{FinalizeExit(i)} transaction is only issued if there is no challenge.

\subsubsection{Sharing Penalization Transactions}

In the OTS-PC scheme with Level 1 privacy, watchtowers do not need their own dispute output of the \txname{CommitExit(i)} transaction (in Figure~\ref{fig:OTS-PC}), because each \txname{CommitExit(i)} transaction accepts only one punishment transaction. Accordingly, Alice's watchtower can use the \txname{CommitPunishBob(i)} transaction.
However, using a separate dispute output allows watchtowers to receive a special reward if they issue valid punishment transactions.

In case of level 2 and level 3 privacy, watchtower dispute outputs do not change with state updates. In these modes,  watchtowers need their own punishment transactions. We justify this by showing how a watchtower could betray a client. Let us assume that Bob tries to unilaterally exit with and old sequence number $j$. Now Alice's watchtower, colluding with Bob, issues a \txname{CommitPunishBob(j)} transaction before Alice can react, but instead of posting the correct one for $j' > j$, it posts one for a sequence number $j' \leq j$. Then, Alice cannot issue the associated \txname{PunishBob(j')} transaction, and the old punishment transaction blocks Alice from punishing Bob.

\begin{itemize}

    \item CommitExit(i): $P_a$/$P_b$: 21 wu. P2WPKH input with aggregate signature: 272 wu. P2WPKH output: 124 wu.
    
    \item AssertExitState(i): Signed sequence number: 800 wu. P2WPKH input with aggregate
    signature: 272 wu. P2WPKH output: 124 wu.
    
    \item FinalizeExit(i): 2 inputs/outputs: 792 wu
    \item Total: 2405 wu
\end{itemize}
\subsection{HTLCs}
\label{htlcs}

We now show how we extend our OTS-PC construction to support routed payments using Hash Time-Locked Contracts (HTLC). HTLCs allow payment channels to be interconnected and to form the Lightning Network. When a transaction output is constrained by an HTLC, the funds can only be spend immediately by one party showing a pre-image of a hash, or by another party only after a timelock. To keep the explanation concise, we present transactions having a single HTLC, though the scheme can support any number of them. 

To create an HTLC-based payment, both parties move to a new state where the locked payment amount is deducted from the sender’s balance but not yet credited to the receiver. Instead, the amount is locked in a new output of \txname{FinalizeExit(i)} that has two specific spending conditions. In case of a payment from Alice to Bob, the HTLC can be spent by Alice after an absolute timelock expires, or by Bob if he has the corresponding pre-image. If it is in the reverse direction, Bob receives the funds after the timelock and Alice with the preimage.

In case an old update state is published and there is a dispute, the funds covered by HTLCs are fully paid to the winner of the dispute. In case of unilateral undisputed exit, the HTLCs remain intact. Figure ~\ref{fig:OTS-PC-HTLC} shows an HTLC output.

\begin{figure}[h]
\centering
\begin{tikzpicture}[x=1cm, y=1cm]
\node[tx, anchor=north, align=center] (ft) at (0,0) {\textit{X}FinalizeExit($i$)};
\node[lbl, anchor=north, align=center, text=gray!65] at (0,-0.75)
  {inputs:\\ Initial\_Funds\_out\\ \textit{X}Ready\_out};

\node[txout, anchor=west] (oa) at (2.6,-0.15) {$A_{BAL}(i)$ to Alice};
\node[txout, anchor=west] (ob) at (2.6,-0.85) {$B_{BAL}(i)$ to Bob};
\node[txout, anchor=west] (h1) at (2.6,-1.75) {AB HTLC};
\node[txout, anchor=west] (h2) at (2.6,-2.85) {BA HTLC};

\node[lbl, anchor=west] (h1a) at (5.0,-1.45) {(1) hash-lock preimage + Bob sig $\rightarrow$ Bob};
\node[lbl, anchor=west] (h1b) at (5.0,-2.05) {(2) absolute timelock + Alice sig $\rightarrow$ Alice};
\node[lbl, anchor=west] (h2a) at (5.0,-2.55) {(1) hash-lock preimage + Alice sig $\rightarrow$ Alice};
\node[lbl, anchor=west] (h2b) at (5.0,-3.15) {(2) absolute timelock + Bob sig $\rightarrow$ Bob};

\draw[bus] (ft.east) -- (1.8,-0.28);
\draw[bus] (1.8,-0.15) -- (1.8,-2.85);
\foreach \n/\y in {oa/-0.15, ob/-0.85, h1/-1.75, h2/-2.85}
  { \draw[fl] (1.8,\y) -- (\n.west); }

\draw[bus] (h1.east) -- (4.3,-1.75);  \draw[bus] (4.3,-1.45) -- (4.3,-2.05);
\draw[fl] (4.3,-1.45) -- (h1a.west);  \draw[fl] (4.3,-2.05) -- (h1b.west);

\draw[bus] (h2.east) -- (4.3,-2.85);  \draw[bus] (4.3,-2.55) -- (4.3,-3.15);
\draw[fl] (4.3,-2.55) -- (h2a.west);  \draw[fl] (4.3,-3.15) -- (h2b.west);
\end{tikzpicture}
\caption{An HTLC output of \txname{FinalizeExit($i$)}. The structure is the same on either exit path, so \textit{X} stands for Alice or Bob.}
\label{fig:OTS-PC-HTLC}
\end{figure}

Informing about active HTLCs to watchtowers is straightforward. Alice can provide her watchtower the list of active HTLCs with their timeout information, together with presigned transactions for spending the HTLCs after timeout, in encrypted form, similar to our level 2 or level 3 watchtower privacy schemes. 

It is worth making explicit why the HTLC amounts must be carried by \txname{FinalizeExit($i$)}, and what that costs. Because an HTLC of a stale state must be seizable by the winner of a dispute, the HTLC outputs cannot become spendable before the dispute window has closed; and because the punishment transactions are pre-signed against \ioname{Initial\_Funds\_out}, whose outpoint is fixed at setup, the funds covering those HTLCs cannot leave that output earlier either. HTLC resolution and dispute resolution are therefore serialised.

This is a consequence of the design choice that gives OTS-PC its $O(1)$ storage. In the Poon-Dryja construction the proof of fraud is a per-commitment revocation secret, which is self-validating and requires no further context: it can be placed first in every output script with no timelock at all, and the BOLT specification uses exactly that property to let an HTLC time out or be fulfilled while the penalty delay is still running. In OTS-PC the proof of fraud is a one-time signature together with evidence that a later state exists, and that evidence is a pre-signed transaction rather than a secret. One one-time key per channel buys constant storage; the price is the loss of a bearer proof of revocation, and with it the ability to decouple HTLC resolution from the dispute window.

\subsection{Compatibility with the Lightning Network}
\label{lncompat}

OTS-PC channels interoperate with the Lightning Network through HTLCs: both constructions lock a payment to the same hash, and that shared lock is what makes a payment crossing the boundary atomic. Neither side needs to understand the other's channel format. What the two do not share is how quickly an HTLC can be resolved on chain, and that difference decides the shape of the integration.

\subsubsection{The forwarding budget}

A Lightning node that forwards a payment must learn the on-chain outcome of its outgoing HTLC before its incoming one expires, and the margin it has is the \texttt{cltv\_expiry\_delta} it advertises. In the Poon-Dryja construction that margin is small, because the HTLC output lives in the commitment transaction and is therefore publishable at once: the node needs one confirmation for the commitment transaction and one for the HTLC transaction. A widely deployed implementation budgets 39 blocks for the pair, assuming 18 blocks as the worst case for a single confirmation. Conventional per-hop deltas are of the same order, around 34 to 40 blocks.

In an OTS-PC the same outcome is four confirmations away rather than two --- \txname{CommitExit}, \txname{AssertExitState}, \txname{FinalizeExit} and the spend of the HTLC output itself --- and the dispute window sits between the second and the third. The delta such a channel must advertise is therefore

$$\Delta \;=\; T \;+\; 4c \;+\; g$$

where $T$ is the dispute window, $c$ the assumed worst case for one confirmation and $g$ a small grace period. With $c=18$ and $g=3$, a channel advertising a conventional delta of 40 blocks cannot serve as a forwarding hop for \emph{any} choice of $T$: the four confirmations alone exceed the budget.

\subsubsection{Sizing the dispute window}

The dispute window is the only timelock left in the construction, and it has one requirement: it must accommodate the punishment, which is the two sequential transactions \txname{CommitPunishX} and \txname{PunishX}. That is the same two-confirmation budget the Lightning implementations assign to their own two-transaction path, so applying the same assumption gives $T \approx 40$ blocks and

$$\Delta \;=\; 40 + 72 + 3 \;=\; 115 \text{ blocks.}$$

Roughly three times the conventional per-hop delta, but well inside the per-hop maximum implementations accept and inside a typical total route budget. An OTS-PC channel can therefore act as a forwarding hop, at the cost of being a less attractive one for pathfinding. Choosing $T$ larger for additional dispute safety raises $\Delta$ one-for-one, and a window of a few hundred blocks takes the channel out of routable range altogether.

\subsubsection{Gateways}

The forwarding budget can be avoided entirely by not forwarding. A gateway node holds an OTS-PC channel on one side and a Lightning channel on the other, and bridges a payment by terminating one leg and originating the other, with both legs locked to the same payment hash.

For a payment arriving from the Lightning Network, the gateway issues the invoice on behalf of the OTS-PC recipient, for a hash whose preimage only the recipient knows. It holds the incoming Lightning HTLC, offers the corresponding HTLC over the OTS-PC channel, and claims the Lightning leg with the preimage the recipient reveals by taking it. In the other direction the gateway accepts the OTS-PC HTLC, re-originates the payment over Lightning, and claims the OTS-PC HTLC with the preimage returned by the Lightning leg. In both directions the gateway can only collect by having paid, so it is never custodial.

Because the gateway is an endpoint of the Lightning payment rather than a hop within it, the delta computed above does not apply to the route: only the gateway's own exposure is bounded by the dispute window, and it can price that risk. The cost is in addressing rather than in safety. An OTS-PC recipient is not independently reachable on the Lightning graph, and in the outbound direction the gateway learns the destination, which is the privacy of a trampoline hop rather than of an onion hop.

A further reason favours this arrangement. A BOLT channel announcement must prove a P2WSH 2-of-2 funding output, whereas an OTS-PC funds into a Taproot covenant address, so OTS-PC channels cannot be announced in the Lightning gossip graph in any case. They are reached through route hints or blinded paths, which is precisely how a gateway advertises the recipients behind it.

\subsubsection{Reducing the budget}

The four confirmations are not reducible, but $T$ enters $\Delta$ only because the HTLC outputs are inside the punishable pot, as explained in section \ref{htlcs}. A construction that caps the value in flight and carves the corresponding amount out of the channel balance at setup would let the HTLC outputs settle on their own schedule while the balances remain subject to the dispute, bringing $\Delta$ close to the Lightning figure. Such a variant introduces a ceiling on in-flight HTLC value, which is not unusual --- Lightning channels already negotiate one --- and a griefing vector bounded by that ceiling. We leave its analysis to future work.

\section{Summary}

In this work, we present a new type of payment channel based on One-Time Signatures (OTS) that enables efficient state revocations and support for lightweight and privacy-preserving watchtower implementations. The payment channel is compatible with the Lightning Network by supporting HTLCs. The channel uses a small set of transactions that are regenerated and signed at each channel state update. Watchtowers only need to store a constant amount of data per monitored channel (the latest punishment transactions), improving upon the standard Lightning Network approach.

\bibliographystyle{IEEEtran}  
\bibliography{references}     

\appendix

\section{Formalization}
\label{formalization}

We formalize our basic construction in Figure \ref{fig:OTS-PC}. We make use of an auxiliary function to create Taproot addresses out of scripts:

\begin{enumerate}
    \item \texttt{TaprootAddress}(list-of-scripts): a P2TR address with a tree containing the given scripts with unspendable NUMS internal key.
    
\end{enumerate}

We make use of an auxiliary functions create verify onetime signatures against a list of public keys: 
\begin{enumerate}  
    \item \texttt{\opc{OT-CSIGV}}(list of OT-public keys): a script that reads from the stack an encoded input message $M$ and one or more OT signatures and verifies the signatures of $M$ against a list of OT public keys. Aborts the script if any of the verifications fail. Leaves the value $M$ on top of the stack. 
\end{enumerate}

We define a transaction as \tuple{list-of-inputs, list-of-outputs}. 
Each input is defined as a prevout structure (\tuple{TxID,index}). However, since we have named each output, we use the name instead. An output is defined as \tuple{amount, address}.
We omit transaction versions and nSequence, and lockTime values for simplicity. One exception must be kept in mind by implementers: \opc{CHECKSEQUENCEVERIFY} is enforced only if the spending transaction has version $\geq 2$ and its input sets nSequence to a relative lock at least as large as the one in the script. A transaction that omits either silently loses the timelock, and with it the dispute window. We additionally use:

\begin{itemize}
\item "-" to indicate that we don't care about this field.

\item \opc{Covenant-Check($i$)} to indicate the signing by all parties involved (Alice and Bob in our case). This can be accomplished by checking 2 signatures using a \opc{CHECKSIGVERIFY} followed by a \opc{CHECKSIG}, or by checking one signature aggregated for both parties using MuSig2 with a single \opc{CHECKSIG}. If present, an index ($i$) indicates which taproot path is being signed. It is written last in every script below, for the reason given at the end of this list.

\item The MuSig2 option requires Taproot. A MuSig2 aggregate is a Schnorr key, and \opc{CHECKSIG} verifies Schnorr signatures only in Taproot; segwit v0 verifies ECDSA over a compressed key and rejects a 32-byte one outright. So an output whose script contains a MuSig2 \opc{Covenant-Check} cannot be P2WSH, and every covenant output below is Taproot for that reason. Taking the P2WSH option and the MuSig2 option together yields an output that can never be spent, and the two are described independently, so nothing in either description looks wrong.

\item \opc{CovAddress} to indicate an address that is an aggregated for Alice and Bob using MuSig2.

\item \opc{CSIG}$(x)$ is defined as the script that pushes the public key $x$, followed by \opc{CHECKSIG}. This is the \emph{terminal} form of a signature check.

\item \opc{CSIGV}$(x)$ is defined as the script that pushes the public key $x$, followed by \opc{CHECKSIGVERIFY}. This is the non-terminal form, for a signature check that is not the last in its script.

\item \opc{CSEQV}$(x)$ is defined as the script that pushes value $x$,  and executes  \opc{CHECKSEQUENCEVERIFY} followed by \opc{DROP}.

\item \opc{CHASHV}$(x)$ is defined as the script that executes \texttt{HASH160}, then pushes the hash value $x$, and executes \texttt{EQUALVERIFY}. Note the order: the opcode hashes the preimage taken from the stack and compares the result against $x$.

\item \opc{CVALV}$(x)$ is defined as the script that pushes the value $x$, and executes \texttt{EQUALVERIFY}.

\item A witness script must finish with exactly one true value on the stack. Each of \opc{CSEQV}, \opc{CHASHV}, \opc{CVALV} and the \opc{LESSTHAN} \opc{VERIFY} of the punishment consumes its operands and pushes nothing, so a script assembled only from them terminates with an \emph{empty} stack, and the output it guards can never be spent. Nothing about that fails noisily: the address is well formed, the spending transaction is well formed, and the defect appears only when a spend is attempted and rejected. Every script below therefore places its signature check last and in the terminal \opc{CSIG} form, so that the value \opc{CHECKSIG} pushes is what satisfies the rule. The ordering is not cosmetic; a reader transcribing these scripts with the checks in any other order will produce unspendable outputs.

\end{itemize}

To facilitate reading, we also embed scripts in outputs as pre-P2SH, while in practice all outputs would be Taproot and specify script hashes, while scripts are revealed in the witness stack in the spending transaction.

We assign names to outputs, so we can refer to the outputs by name instead of by transaction id/index.

We omit the definitions of WTA and WTB disputes since they are similar to Alice and Bob disputes respectively, only paying a small fee to a  watchtower address out of the total funds paid.

The balance $I_{BAL}$ represents the total balance deposited in the payment channel. Balances $A_{BAL}(i)$ and $B_{BAL}(i)$ represents the balances of Alice and Bob after state update $i$, not including the amounts that may be locked in HTLCs nor the amounts committed to be paid in transaction fees.

We define $\epsilon$ as the amount in bitcoins that is the minimum necessary to make the largest transaction in our graph standard. The standard transaction rules dictate that a certain fee-rate must be paid by transaction inputs, even when child transactions are paying for parent fees using CPFP. The \txname{Setup} transaction moves $k\epsilon$ bitcoins to each exit output, where $k$ is the worst-case depth of transactions below it. In the basic OTS-PC construction of this appendix the worst case is $k=3$ for an exit output (\txname{CommitExit}, \txname{AssertExitState}, \txname{FinalizeExit}) and $k=2$ for a dispute output (\txname{CommitPunishX}, \txname{PunishX}). The privacy variants of section \ref{watchtowers} insert a transaction into one of these chains and raise the corresponding $k$ by one; the budget each requires is given there.

The scheme uses 3 addresses defined during setup: AliceAddress, BobAddress and SharedAddress. AliceAddress and BobAddress are the addresses where Alice and Bob will receive the funds after unilateral exit. The SharedAddress is used only to let Alice or Bob bump transaction fees using the Child-Pay-For-Parent (CPFP) method. The only transactions that need this output are \txname{AliceAssertExitState} and \txname{BobAssertExitState}. The remainder transactions can be accelerated by Alice or Bob using the package relay method. This requires creating a new transaction \txname{Forward} to transfer the newly received (but unconfirmed) bitcoins to a new owned address, paying higher-than-normal transaction fees. With package relay nodes/miners accept and relay packages of related transactions instead of each transaction in isolation. In our case, it makes miners include all (unconfirmed) parent transactions of \txname{Forward}.

Each owner has its own exit path, rooted at its own \ioname{Alice\_Exit\_out} or \ioname{Bob\_Exit\_out} output. Both paths are defined symmetrically, so below we give Alice's path in full and only note Bob's where it differs. Although every transaction is signed by both owners, the signature on a path's first transaction is delivered in one direction only: Bob gives Alice his signature on \txname{AliceCommitExit($i$)} and never receives hers, and vice versa. Each owner can therefore publish its own path and neither can publish the other's.

\txname{Setup} = \tuple{
    \tuple{ \ioname{Funds\_From\_Alice\_in}, \ioname{Funds\_From\_Bob\_in}},
    \tuple{ \ioname{Initial\_Funds\_out}, \ioname{Alice\_Exit\_out}, \ioname{Bob\_Exit\_out},
    \ioname{Bob\_Disputes\_out}, \ioname{Alice\_Disputes\_out}, 
    \ioname{WTB\_Disputes\_out}, \ioname{WTA\_Disputes\_out} }
}

\ioname{Funds\_From\_Alice\_in} =\tuple{-, -}

\ioname{Funds\_From\_Bob\_in}   =\tuple{-, -} 

\ioname{Initial\_Funds\_out} = \tuple{ $I_{BAL}$ ,  \opc{CovAddress} }

\ioname{Alice\_Exit\_out} = \tuple{ $3\epsilon$, \opc{CovAddress} } 

\ioname{Bob\_Exit\_out} = \tuple{ $3\epsilon$, \opc{CovAddress} } 

\ioname{Bob\_Disputes\_out} = \tuple{ $2\epsilon$, \opc{CovAddress} } 

\ioname{Alice\_Disputes\_out} = \tuple{ $2\epsilon$, \opc{CovAddress} } 

\ioname{WTB\_Disputes\_out} = \tuple{ $2\epsilon$,  \opc{CovAddress} } 

\ioname{WTA\_Disputes\_out} = \tuple{ $2\epsilon$,  \opc{CovAddress} } 
 
\txname{AliceCommitExit($i$)} = \tuple{
    \tuple{ \ioname{Alice\_Exit\_out}} ,
    \tuple{ \ioname{AliceCommitExit\_out}}
}

\ioname{AliceCommitExit\_out} = \tuple{ $2\epsilon$, 
 \texttt{TaprootAddress}( \tuple{ 
      \opc{OT-CSIGV}( $K^A_{pub}$ ) \opc{CVALV($i$)}
      \opc{Covenant-Check}
      })
 }

\txname{BobCommitExit($i$)} and \ioname{BobCommitExit\_out} are identical with \ioname{Bob\_Exit\_out} and $K^B_{pub}$ in place of \ioname{Alice\_Exit\_out} and $K^A_{pub}$.

\txname{AliceAssertExitState($i$)} = \tuple{
    \tuple{ \ioname{AliceCommitExit\_out}} ,
    \tuple{ \ioname{AliceReady\_out},\ioname{AliceCPFP\_out}}
}

\ioname{AliceReady\_out} = \tuple{ $\epsilon$, {
 \texttt{TaprootAddress}(\tuple{ 
 \tuple{ 
 \textbf{[1]} 
  \opc{CSEQV}( $T$ ) 
  \opc{Covenant-Check(1)} },
 \tuple{ 
 \textbf{[2]} 
   \opc{CHASHV($H_a$)} 
   \opc{CHASHV($H_b$)} 
   \opc{Covenant-Check(2)} 
  }
 })
 }
}

\ioname{AliceCPFP\_out}= \tuple{ $\epsilon$, SharedAddress }

\txname{AliceFinalizeExit($i$)} = \tuple{
    \tuple{ \ioname{Initial\_Funds\_out}, \ioname{AliceReady\_out}},
    \tuple{
        \tuple{ $A_{BAL}(i)$, AliceAddress }
        \tuple{ $B_{BAL}(i)$, BobAddress }
    }
}

\txname{BobAssertExitState($i$)}, \ioname{BobReady\_out}, \ioname{BobCPFP\_out} and \txname{BobFinalizeExit($i$)} mirror the above. Note that \txname{AliceFinalizeExit($i$)} and \txname{BobFinalizeExit($i$)} pay the same amounts to the same addresses, and both spend \ioname{Initial\_Funds\_out}: only one of them can ever confirm, and it does not matter which.

\txname{CommitPunishAlice($i$)} = \tuple{
    \tuple{ \ioname{Bob\_Disputes\_out}},
    \tuple{ \ioname{PunishAlice\_out} }
}
    
\ioname{PunishAlice\_out} = \tuple{ $\epsilon$, \texttt{TaprootAddress}({
          \tuple{ 
           \opc{OT-CSIGV}( $K^A_{pub}$ )  
           \tuple{i} 
           \opc{LESSTHAN} 
           \opc{VERIFY} 
           \opc{Covenant-Check} } 
        }
}

\txname{PunishAlice($i$)} = \tuple{
    \tuple{ \ioname{Initial\_Funds\_out}, \ioname{PunishAlice\_out}},
    \tuple{
        \tuple{ $I_{BAL}$, BobAddress }
    }
}

\txname{CommitPunishBob($i$)} = \tuple{
    \tuple{ \ioname{Alice\_Disputes\_out}},
    \tuple{ \ioname{PunishBob\_out} }
}
    
\ioname{PunishBob\_out} = \tuple{ $\epsilon$, \texttt{TaprootAddress}({
          \tuple{ 
           \opc{OT-CSIGV}( $K^B_{pub}$ )  
           \tuple{i} 
           \opc{LESSTHAN} 
           \opc{VERIFY} 
           \opc{Covenant-Check} } 
        }
}

\txname{PunishBob($i$)} = \tuple{
    \tuple{ \ioname{Initial\_Funds\_out}, \ioname{PunishBob\_out}},
    \tuple{
        \tuple{ $I_{BAL}$, AliceAddress }
    }
}


\end{document}